# Conformations of Macromolecules and their Complexes from Heterogeneous Datasets

P. Schwander, R. Fung, and A. Ourmazd

Dept. of Physics, University of Wisconsin Milwaukee, 1900 E. Kenwood Blvd, Milwaukee, WI 53211

*Corresponding author:* A. Ourmazd; Tel: (414) 229-2610; Email: Ourmazd@uwm.edu

Total words: 5,226

**Abstract**

We describe a new generation of algorithms capable of mapping the structure and conformations of macromolecules and their complexes from large ensembles of heterogeneous snapshots, and demonstrate the feasibility of determining both discrete and continuous macromolecular conformational spectra. These algorithms naturally incorporate conformational heterogeneity without resort to sorting and classification, or prior knowledge of the type of heterogeneity present. They are applicable to single-particle diffraction and image datasets produced by X-ray lasers and cryo-electron microscopy, respectively, and particularly suitable for systems not easily amenable to purification or crystallization.

(88 words)

# 1. Introduction

Biomolecular interactions, through the formation of transient or robust complexes, are at the center of cellular function and life itself. There is increasing recognition that biological macromolecular complexes exist in a range of conformations, and that these can play a vital role in their function. The virulence of the dengue virus, for example, strongly depends on transitions in its protein contacts and conformational rearrangements [1]. A deep understanding of the nature and role of conformational variety in biological function would revolutionize our knowledge of key processes ranging from basic cell function to pathological states.

Despite powerful contributions to the study of proteins and some complexes, X-ray crystallography and NMR have limitations. With notable exceptions, the constraints imposed by crystals have limited the role of X-ray crystallography in elucidating conformational variety. NMR, while able to study conformations in biomolecules of modest size, has not been extensively applied to larger systems. Cryo-electron microscopy (cryo-EM), fortified with increasingly sophisticated algorithmic approaches [2-4], has been extensively used to study macromolecular complexes. However, conformational variety presents a challenge to cryo-EM methods, which often assume its absence. When conformational variety has been explicitly addressed, the results, won with effort and ingenuity, have provided tantalizing evidence of a rich variety of conformations, even in well-studied systems [4-6]. The difficulties faced in addressing macromolecular complexes and interactions by otherwise successful structural means have led to the recognition that dealing with heterogeneity represents an important challenge in need of urgent attention. For example, there is currently no direct means for mapping the continuum of three-dimensional (3D) structures assumed by conformationally flexible complexes, such as the therapeutically important G protein-coupled receptors (GPRCs) [6].

Heterogeneity is pervasive, because the observation of an ensemble of macromolecules in reaction or interaction naturally leads to a collection of snapshots from non-identical objects. The ability to extract structural information from large datasets of snapshots obtained from configurationally and conformationally heterogeneous ensembles of complexes would substantially advance our understanding of macromolecular conformations and their role in biology. However, structure recovery methods such as cryo-EM [7, 8] and emerging X-ray Free Electron Laser (XFEL) "diffract-then-destroy" approaches [9-13] are generally predicated on viewing a series of identical objects from different angles. The obvious method of "sorting" the data into classes, each stemming from nominally identical objects is fraught with difficulty: the number and types of classes are often unknown; sorting must be performed at very low signal-to-noise ratios ("SNR" $\leq 0.1$); and residual heterogeneities persist, even when the classes are small. Indeed, there is evidence that the resolution achieved by cryo-EM is often limited by sample heterogeneity.

Heterogeneity can be tackled by sorting with reference to templates, which often can only be guessed at. The dangers in this approach are well known [8]. For example, the image of any individual, say Einstein or Newton, can emerge from random noise, depending on whose portrait was used as a template. Approaches based on Bayesian inference and maximum likelihood are powerful (see, e.g., [4]), but inherently favor the discovery of discrete conformations. Their computational expense and scaling behavior also limit their practical application to a small number of conformations. In short, the analysis of existing cryo-EM datasets and those emerging from XFEL techniques is severely hampered by the absence of algorithms able to deal naturally and efficiently with heterogeneity.

Methods recently developed in our group offer the possibility to extract structural and conformational information from heterogeneous datasets directly, and efficiently [14-18]. These methods combine techniques from Riemannian geometry, graph-theoretical dimensionality reduction ("manifold embedding"), and scattering physics. Fundamentally, however, they are based on the simple recognition that each snapshot in a heterogeneous dataset provides information about all states of the system under observation. For example, the view from the back of a person's head has valuable information about the full-frontal view, because it reveals where the ears are – irrespective of whether the person is smiling or not. Thus, the entire dataset can be used to reconstruct each state of the system, even when the dataset is heterogeneous. This approach substantially increases the available information, allowing one to operate at significantly lower signal levels than needed today. And it offers the possibility to use the information from all conformations to recover the 3D structure of each.

Here, we describe our approach and present results demonstrating its capabilities in the context of simulated XFEL diffraction and cryo-EM image snapshots. Section 2 outlines the approach in conceptual terms. Mathematical underpinning can be found in [17, 18] and the references therein. Section 3 presents results on simulated diffraction snapshots of an enzyme undergoing large conformational changes. More subtle changes are likely to remain beyond the reach of XFEL experiments for some time. Section 4 describes results on simulated ultralow signal cryo-EM snapshots of objects undergoing more subtle conformational changes, imaged in the presence of large systematic changes such as defocus variation. These results pertain to structures often used to benchmark different approaches, in order to facilitate comparative assessment. We discuss our results and future challenges in section 5, and summarize and conclude the paper in Section 6.

## 2. Conceptual Outline of Approach

Our approach is able to recover 3D structural and conformational information from ultralow signal, heterogeneous datasets without templates or pre-classification. This stems from the recognition that: a) datasets from ensembles of macromolecules in reaction or interaction are necessarily heterogeneous; and b) the information content of the entire dataset can be used to recover the 3D structure corresponding to each conformation of the system. Rather than avoiding heterogeneity by careful experimental means or through sorting the snapshots into nominally identical classes, the approach exploits heterogeneity to increase the available information substantially.

A more technical, but equally important aspect rests on the recent discovery that snapshots produced by scattering experiments reside on data manifolds with specific symmetries [17]. These symmetries stem from the nature of operations in space, rather than the object itself, and are thus entirely general. This allows one to project complex, noisy datasets on known manifolds, much as one fits data to a polynomial of known type. Because the fit is determined by the entire dataset, this represents an efficient and noise-robust means of extracting information. In principle, multiple species result in multiple manifolds, and the properties of each manifold ("coefficients of each fit") can be used to deduce the 3D structure and conformational continuum of each species. Slightly more technically, modern graph-theoretic manifold embedding techniques [16, 19-27] can be used to find manifolds produced by scattering. Similar manifolds have been previously encountered in certain general relativistic models of the universe, and are thus well known [28, 29]. Laplacian eigenfunctions of manifolds produced by scattering can be deduced from such models, and used to extract structural and conformational information from scattering data [17, 18]. Our algorithms are noise-robust, computationally efficient, work with existing and emerging large datasets comprising up to 20 million snapshots, and can be incorporated into existing structure recovery platforms for enhanced reach and impact.

Our approach can be simply understood by considering, for example, a particle with three orientational degrees of freedom. As the particle orientation is changed, the changes in the pixel intensities are a function of only three parameters. This imposes a strong correlation among the pixel intensities, which can be used to determine the snapshot orientations, and thus determine the 3D structure [16, 30-32]. Specifically, a snapshot consisting of $p$ pixels can be represented as a $p$-dimensional vector, with each component representing the intensity value at a pixel. The fact that the intensities are a function of only three parameters means that the $p$-dimensional vector tips all lie on a 3D hypersurface ("manifold") in the *p*-dimensional space of intensities (Fig. 1). This manifold is an expression of the correlated way in which the pixel intensities change with the particle orientation. In fact, each point on the manifold represents a snapshot from a specific object orientation.

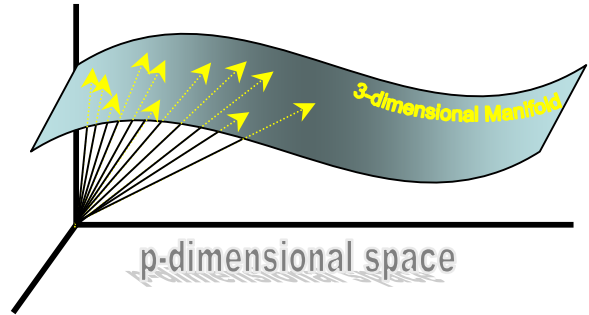

**Figure 1.** Manifold as expression of correlation. An object has only three orientational degrees of freedom. This means that the *p* pixel intensities in a snapshot change in a correlated fashion with object orientation. This correlation is described by a 3D manifold in the *p*-dimensional space of pixel intensities.

Discovering the manifold in a noisy dataset constitutes the first step in the analysis. Starting with the Euclidean distance between vectors representing snapshots in the *p*-dimensional data space, a number of powerful graph-theoretic techniques can be used to discover low-dimensional manifolds underlying the high-dimensional data (see, e.g., [19, 23-26, 33-35]). Each algorithm has its strengths and limitations, with the most common problem being noise sensitivity [36]. Our approach incorporates three different manifold embedding approaches based on Generative Topographic Mapping (GTM) [19, 20], Isomap [26], and Diffusion Map [25, 33, 37]. In each case, extensive effort was required to achieve noise-robustness. The resulting algorithms can operate at signal levels as low as -20dB (1/100 on a linear scale), depending on the application [32].

Once the data manifold has been determined and embedded in a suitable space, one must discover how to interpret the outcome. Specifically, one needs to identify the operations connecting any two points on the manifold. For example, in order to reconstruct a 3D image of a given conformation, one must identify all points on the manifold which can be reached by SO(3) operations (3D rotations) alone. Similarly, to map conformations, all points on the manifold connected by conformational operations alone must be identified. This constitutes the second important step in our approach.

Manifolds are best described in differential geometrical terms, with the metric – the local measure of distance – playing an important role. Using a differential geometric formulation of scattering, we have been able to relate changes in the data manifold to specific operations [18]. In non-technical terms, one would like to relate infinitesimal changes in the intensity distribution in a snapshot to the corresponding infinitesimal operations affecting the orientation and/or conformation of the object. In other words, one would like to relate the metric of the data manifold to the metric of the manifold of operations. This would allow one to determine the rotation and/or conformation operations connecting any pair of snapshots. Achieving this is tantamount to having a model of the object, in the sense that, given any snapshot, any other corresponding to a desired object orientation and conformation can be produced on demand. The problem, however, is that the metric of data manifolds produced by scattering is not simply related to that of the manifold of operations. For cases involving orientational changes only, we have solved this problem in two steps. First, we have shown that the metric of data manifolds produced by scattering onto a 2D detector can be decomposed into two parts, one with high symmetry, plus an object-specific "residual" with low symmetry [17]. Second, using results from general relativity and quantum mechanics, we have shown that the (Laplace-Beltrami) eigenfunctions of the high-symmetry part are directly related to those of the manifold of rotation operations under a wide range of scattering conditions [17]. This allows one to deduce the orientation corresponding to each snapshot. Fig. 2 demonstrates 3D structure recovery by this approach down to very low signal levels [18].

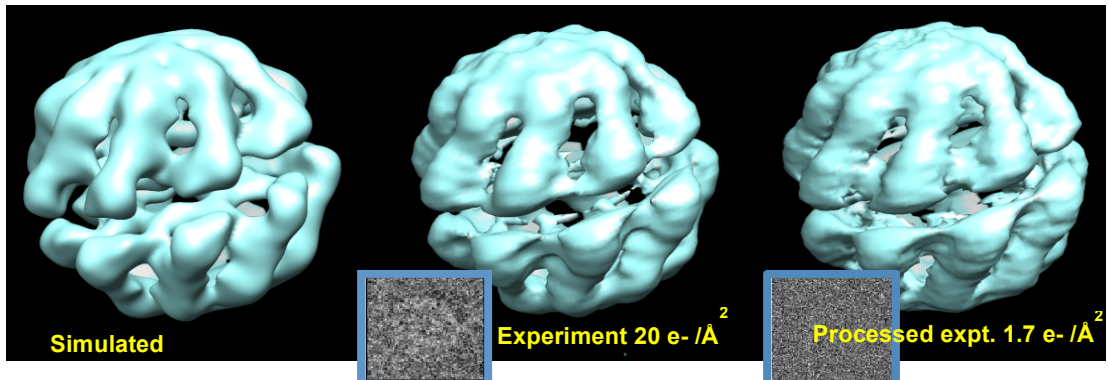

**Figure 2.** 3D reconstructions of chaperonin molecule from cryo-EM snapshots [18]. Left: Noise-free simulation. Center: With experimental images at 20 electrons/Å$^2$. Right: With images obtained by processing experimental snapshots to approximate a dose of 1.7 electrons/Å$^2$. Insets show typical snapshots at each electron dose.

## 3. Conformations from Diffraction Snapshots

We have previously shown that experimental XFEL diffraction snapshots stemming from an unknown mixture of species can be sorted with high accuracy [15]. This potentially offers a post-processing route to mitigating the solution purification problem. Here, we are concerned with determining conformations of a single macromolecule or macromolecular assembly. Experimental single-particle XFEL snapshots are currently dominated by extraneous effects, such as stochastic variations in the beam intensity and inclination, the diameter and position of the liquid jet containing the particles, and detector saturation and nonlinearity. These overwhelm the signal from the particle itself, and cannot be alleviated, for example, by increasing the incident beam intensity.

Experience in cryo-EM, however, has shown that such effects can be alleviated by advances in experimental and algorithmic techniques. Until improved datasets are generally available, algorithm development must rely on simulated snapshots.

### 3.1. Discrete Conformations

We have previously shown that when simulated single-particle XFEL snapshots emanate from different discrete conformations of the same complex, our approach automatically sorts the diffraction snapshots into separate conformational classes and determines their orientations [14]. Fig. 3 shows the results of a manifold embedding analysis, when a mixture of randomly oriented diffraction snapshots from the closed and open conformations of the enzyme adenylate kinase (ADK, PDB identifier: 1ank and 4ake, respectively) were presented to the algorithm at the signal level corresponding to 0.04 photons/Shannon pixel at 0.18nm with shot noise [14]. Because of their chemical identity, the conformations of ADK are difficult to separate. As shown in Fig. 3, our algorithm automatically sorts the snapshots into different manifolds, and determines the orientations of the members of each set of snapshots. No prior information was provided to the algorithm regarding the type or number of conformations present.

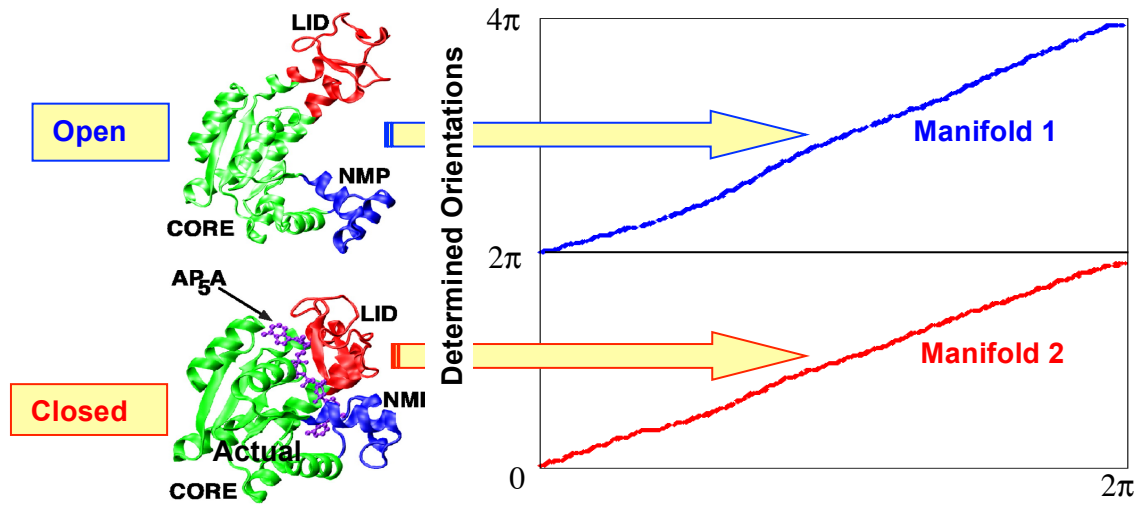

**Figure 3.** Sorting snapshots from different conformations. A mixture of simulated diffraction snapshots from the molecule ADK in its open and closed conformations at 0.04 photons/pixel at 0.18nm with shot noise is automatically sorted into different manifolds and the orientation of each snapshot determined [14].

The confidence with which sorting was performed can be deduced as follows. Noise causes the vectors representing the snapshots not to lie exactly on the manifolds, thus imparting a certain "thickness" to each manifold. This can be quantified in terms of the widths (standard deviations) of the distributions of vectors about the manifolds. At the signal level of 0.04 photon/pixel with shot noise, the smallest separation between the two manifolds is ~10 standard deviations. This means that diffraction snapshots from the different conformations are sorted with extreme fidelity, even in the presence of substantial noise. This level of confidence clearly cannot be expected with experimental data. Nevertheless, results obtained with simulated data provide an indication of the efficiency with which different conformations of a molecule may be identified and separated. We note that larger objects, such as macromolecular complexes, produce larger signals, and should thus be more readily amenable to our approach.

**3.2. Continuous Conformations**

Macromolecular complexes are even more likely to possess conformational continua than discrete conformations. However, one must walk before learning to run. We have therefore used the unfolding of ADK to demonstrate the principle of mapping conformational continua. The unfolding process of ADK was simulated by molecular dynamics as follows. The coordinates of ADK from *E. coli* in the open state (Protein Data Bank identifier: 4ake) were placed in a spherical droplet of water and simulated at a nominal temperature of 850 K using NAMD [38]. 12,500 diffraction snapshots were simulated from 100 conformations, with each conformation assuming 125 orientations about one axis. Snapshots were provided to a modified version of the Isomap manifold embedding algorithm [26], and the resulting manifold displayed through its projections along the first three principle components (Fig. 4). It is clear that orientational and conformational variations give rise to a tubular manifold. Qualitatively, the closed cross-sections of the tube represent orientational change, while paths terminating at the tube ends include conformational change. It can be readily shown that the manifold is Riemannian. Due to the SO(3) symmetry operations involving molecular orientation, the manifold has SO(3) symmetry in some directions. Such manifold have received considerable attention in general relativity [28, 29]. We have demonstrated that these techniques, suitably modified, can be used to recover the 3D structure of biological objects with computational complexities $10^4$ times higher than previously possible [17]. Following general relativistic models for the evolution of the universe [29], we are exploring extensions of this approach to mapping conformational continua [A. Ourmazd et al., unpublished]. Fig. 5, for example, shows projection on the first Diffusion Map eigenfunction of the manifold produced by snapshots of melting ADK free to assume any orientation in 3D. From the molecular dynamics simulation described above, 2.4 million diffraction snapshots were simulated from 12 conformations separated by equal time steps of 37.5ps, with each conformation assuming 200,000 orientations. The snapshots were then analyzed with the Diffusion Map algorithm. As depicted in Fig. 5 (a) the histogram clearly distinguishes all 12 conformations except the last two, which overlap.

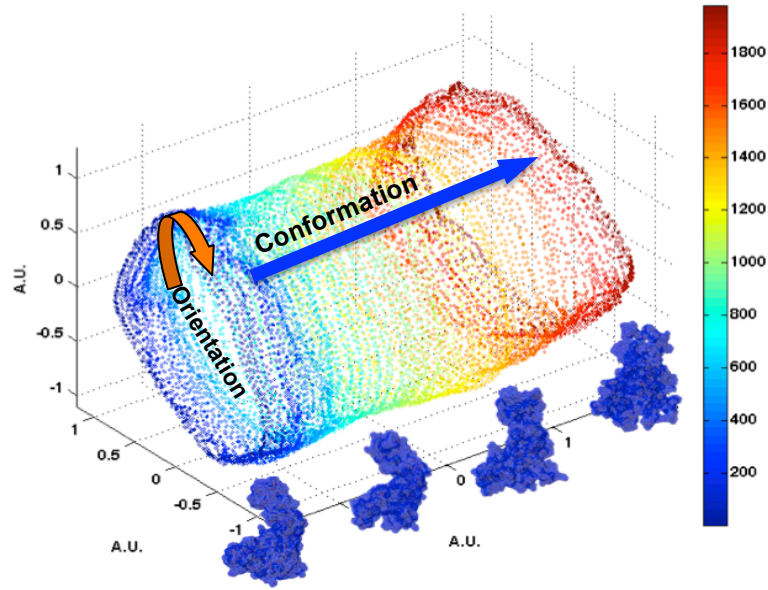

**Figure 4.** Manifold traced out by 100 conformations of an unfolding ADK molecule, each able to assume one of 125 orientations about a single axis. The molecular conformations corresponding to different points on the manifold are shown in the insets.

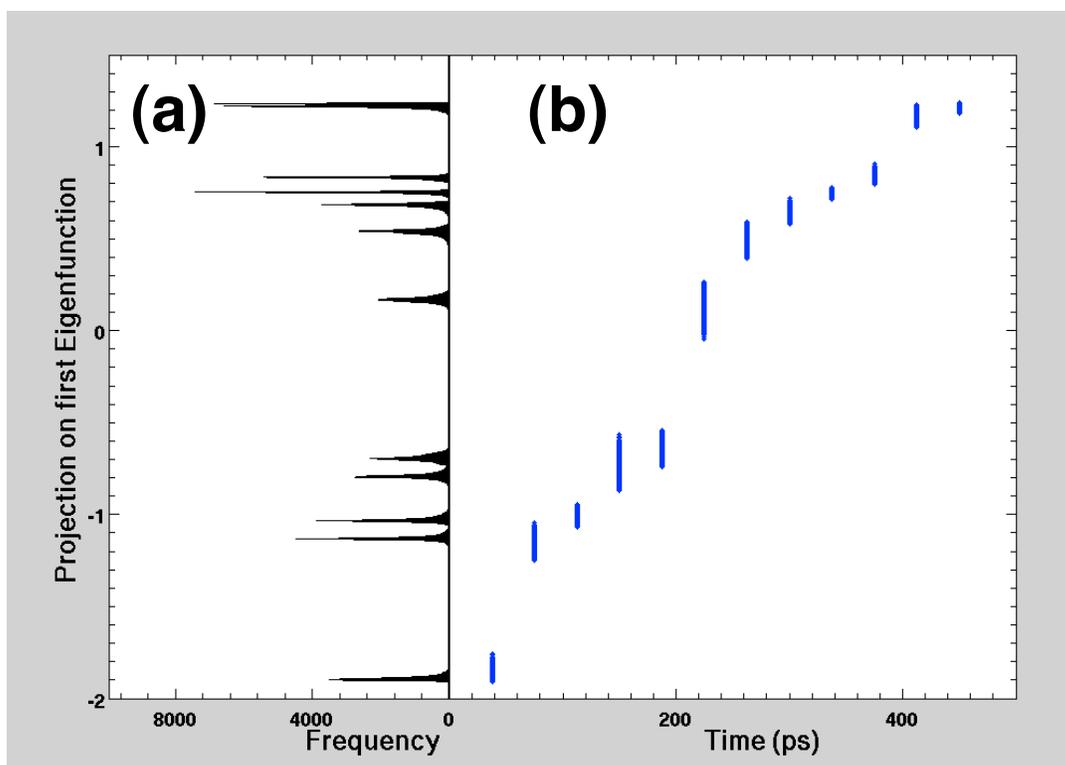

**Figure 5.** Eigenfunction revealing conformational information. The manifold was formed by diffraction snapshots of 12 conformations of melting ADK free to assume any orientation in 3D. Projections on the first eigenfunction are shown as a histogram (a), and as a function of time (b).

Thus the first eigenvector directly provides conformational information, confirming the possibility to identify directions of orientational and conformational change on the manifold. These results offer a potentially promising route to using heterogeneous XFEL datasets to map conformational continua in macromolecular complexes. The possibility to use single-particle techniques to map conformational continua is new, and expected to remove a major bottleneck in the study of complexes, where structural flexibility can play a prominent role.

### 4. Conformations from Cryo-EM Images

Conformational changes are, as a rule, far more subtle than those displayed by an unfolding macromolecule. In such cases, the effect of even small orientational changes can overwhelm the signal due to typical conformational variations. Under these circumstances, a different approach is needed to map conformations.

We now outline a manifold-based approach capable of sorting with high fidelity, simulated noisy single-particle cryo-EM snapshots of mildly heterogeneous particles, and demonstrate this capability in the context of ribosome complexes with and without growth elongation factor (EFG) (Fig. 6). In order to facilitate comparison with the results obtained by other approaches, we utilize a dataset often used for benchmarking the performance of conformational sorting algorithms [39]. A mixture of 200,000 snapshots from ribosome able to assume any orientation in 3D with and without EFG was simulated at defocus values ranging from -1.5 to -2.5$\mu$m with added background and shot noise corresponding to an SNR of -12dB (0.06 on a linear scale). These parameters are typical of experimental cryo-EM snapshots.

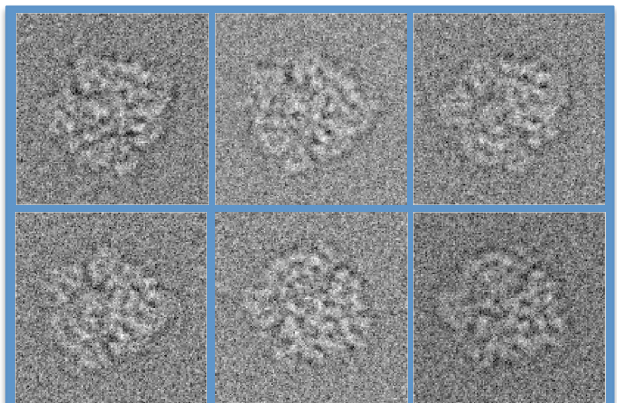

**Figure 6.** Variance-normalized, simulated noisy cryo-EM images of ribosome with and without EFG in different orientations, in random order.

The analysis proceeds as follows. First, the snapshot orientations are determined irrespective of the (unknown) conformational states of the particles. This is possible, because the effect of orientational change dominates. Any algorithm capable of determining orientation can be used for this purpose. Both standard cryo-EM [8] and manifold-based approaches are able to determine orientation with an accuracy of about one Shannon angle. Conformational discrimination is achieved by means of a special kernel for the Diffusion Map algorithm to extract the small conformational signal in the presence of large changes due to orientation, viz.

$$W_{ij} = \begin{cases} \exp\left(-\dfrac{D_{ij}^2}{\sigma^2}\right) & \Delta\theta_{ij} < \theta_\varepsilon \\ 0 & \text{otherwise} \end{cases},$$

where $W_{ij}$ is the weighting factor for a pair of snapshots $i$ and $j$ separated by a great-circle angular distance $\Delta\theta_{ij}$ and a modified Euclidean distance $D_{ij}$ (to be defined below), $\theta_\varepsilon$ an upper bound for $\Delta\theta_{ij}$, and $\sigma$ the Gaussian kernel width. At a sufficiently small $\theta_\varepsilon$ the conformational signal dominates. By assigning zero weight to snapshot pairs separated by more than $\theta_\varepsilon$, this weighting scheme is primarily sensitive to changes in conformation only. To enhance this further, we retain only a small number of the shortest distances in the Diffusion Map analysis.

Since cryo-EM snapshots can also differ by the defocus at which they were obtained, the effect of the defocus on two otherwise identical snapshots must be eliminated. This is achieved by the following definition of the Euclidean distance $D_{ij}$:

$$D_{ij} = \sum_p \left| \text{PSF}_j \otimes I_i - \text{PSF}_i \otimes I_j \right|^2$$
$$= \sum_p \left| \text{PSF}_j \otimes (\text{PSF}_i \otimes P_i) - \text{PSF}_i \otimes (\text{PSF}_j \otimes P_j) \right|^2 \quad .$$
$$= \sum_p \left| \text{PSF}_i \otimes \text{PSF}_j \otimes \{P_i - P_j\} \right|^2$$

For each snapshot $i$, $I_i$ represents the image intensity distribution, $\text{PSF}_i$ the microscope point-spread function and $P_i$ the projected potential, and $\otimes$ the convolution operator. This "double-filtering" scheme ensures a zero Euclidean distance between two snapshots stemming from the same projected potential, but differing in defocus values. For computational efficiency, the distances are calculated in Fourier space so that convolution becomes multiplication. With $\tilde{I}_i$ the Fourier transform of the image and $\text{CTF}_i$ the Fourier transform of the point-spread function, application of Parseval's theorem [40] yields

$$D_{ij} = \sum_q \left| \text{CTF}_j \cdot \tilde{I}_i - \text{CTF}_i \cdot \tilde{I}_j \right|^2 \quad .$$

Fig. 7 shows the results obtained by appropriate embedding of ribosome with and without EFG, as outlined above, with following parameters: $\theta_\varepsilon = 0.08$ corresponding to two Shannon angles, the number of nearest neighbors retained =3, and $\sigma$ determined as described in [41]. In this plot, each snapshot is represented by its coordinate in the plane defined by the second and third Diffusion Map eigenvectors, with the snapshots colored red and blue corresponding to ribosome with and without EFG, respectively. The cutting line (black) separates the two clusters with a sorting fidelity of 99.96%.

The table below shows a compilation of the results at constant and varying defocus, and different means of obtaining orientational information, in comparison with benchmarks from the literature [39, 42, 43].

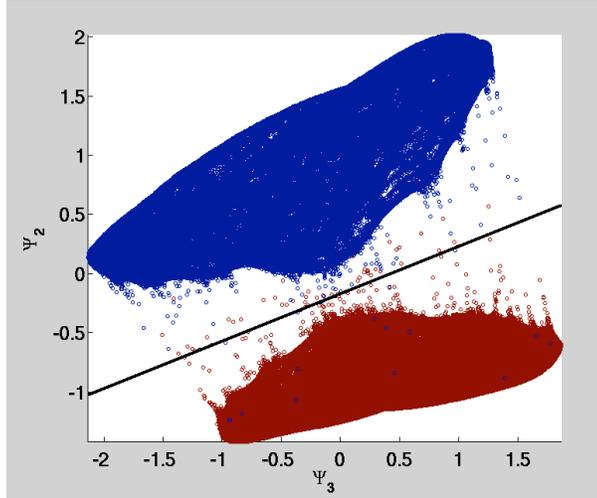

**Figure 7.** Two ribosome conformations (with and without EFG) are separated with 99.96% accuracy.

| Defocus (-ve μm) | Orientational Accuracy ($\theta_{Shannon}$) | Sorting Fidelity | Benchmarks | Remarks |
|---|---|---|---|---|
| 2 | ~1 | 100% | | Known orientations |
| 2 | 2 | 99.9995% | 99.7% | Diff Map orientations |
| 1.5 – 2.5 | ~1 | 99.97% | | Known orientations |
| 1.5 – 2.5 | ~1 | 99.96% | 87% | SPIDER orientations |

In summary, these results show that the two conformations can be identified with 99.96% accuracy in the presence of experimental noise and defocus variations, compared with the previously best published fidelity of 87% [39].

## 5. Discussion

The techniques most commonly used to investigate conformations of macromolecules implicitly assume the presence of discrete conformations, sometimes requiring starting "templates" for each conformation, or at least some knowledge of the number of conformations present. The investigation of conformational continua, perhaps one of the most important aspects of molecular function, has remained difficult, if not beyond reach. The approach we have outlined offers a natural means of dealing with conformations, whether discrete or continuous, without bias or a priori knowledge. Both diffraction and image snapshots are amenable to this approach, even in the presence of overwhelming noise and (known) systematic variations such as defocus.

The majority of results presented here, however, pertain to simulated snapshots, and successful experimental demonstration remains an important future task. In the case of XFEL snapshots, this must include means of dealing with the effects of unknown stochastic variations in the intensity, position, and inclination of the incident beam, and the geometrical parameters characterizing the way the particle was illuminated in each shot.

## 6. Summary and Conclusions

We have described an approach naturally suited to investigating macromolecular conformations and conformational continua using heterogeneous sets of diffraction or image snapshots, without the need for prior assumptions regarding the nature of the conformational variety present. Simulations show the approach to be capable of operating with extreme fidelity at signal-to-noise levels typical of experimental data, at least in the case of cryo-EM snapshots. This offers a promising route to investigating conformational variety in macromolecular systems and its role in biological function.

**Acknowledgments**

We are grateful to D. Giannakis for discussions and some manifold calculations, G. N. Phillips for discussions and molecular dynamics simulations, J. Frank and H. Y. Liao for discussions, simulation of ribosome snapshots, and assistance with cryo-tomographic reconstruction, and W. Chiu and J. Zhang for experimental cryo-EM data. This research was supported by the US Dept. of Energy, Office of Science, Basic Energy Sciences under award DE-FG02-09ER16114. The publication of this work was supported by the US National Science Foundation under award STC 1231306.